\documentclass[5p,number,sort&compress]{elsarticle}                            

\usepackage[utf8]{inputenc}
\usepackage[T1]{fontenc}
\usepackage{amsmath}
\usepackage{amssymb}
\usepackage{textcomp}
\usepackage[caption=false]{subfig}
\usepackage{booktabs}

\newcommand{\un}[1]{\,{\rm #1}}

\newcommand{\dd}{\mathrm{d}}
\newcommand{\degree}{^\circ}
\newcommand{\QQ}{{\cal Q}}
\newcommand{\UU}{{\cal U}}
\DeclareMathOperator{\cov}{Cov}

\DeclareMathOperator{\RMS}{RMS}

\usepackage[usenames]{color}
\definecolor{dark-red}{RGB}{238,0,3}
\definecolor{dark-blue}{RGB}{0,0,165}
\definecolor{dark-purple}{RGB}{224,39,215}
\definecolor{dark-green}{RGB}{12,109,0}

\usepackage{figurewidth}
\setfigurewidth{0.6}{1}

\journal{Astroparticle Physics}

\begin{document}

\begin{frontmatter}

\title{Analyzing the Data from X-ray Polarimeters with Stokes Parameters}
\makeatletter
\def\@author#1{\g@addto@macro\elsauthors{\normalsize%
    \def\baselinestretch{1}%
    \upshape\authorsep#1\unskip\textsuperscript{%
      \ifx\@fnmark\@empty\else\unskip\sep\@fnmark\let\sep=,\fi
      \ifx\@corref\@empty\else\unskip\sep\@corref\let\sep=,\fi
      }%
    \def\authorsep{\unskip,\space}%
    \global\let\@fnmark\@empty
    \global\let\@corref\@empty  
    \global\let\sep\@empty}%
    \@eadauthor={#1}
}
\makeatother

\author{F.~Kislat\corref{cor}}
\ead{fkislat@physics.wustl.edu}
\author{B.~Clark}
\author{M.~Beilicke}
\author{H.~Krawczynski}
\address{Washington University in St. Louis, Department of Physics and McDonnell Center for the Space Sciences, One Brookings Dr., CB 1105, St.\ Louis, MO 63130, United States}
\cortext[cor]{Corresponding author}

\begin{abstract}
X-ray polarimetry promises to deliver unique information about the geometry of the inner accretion flow of astrophysical black holes and the nature of matter and electromagnetism in and around neutron stars. 
In this paper, we discuss the possibility to use Stokes parameters -- a commonly used tool in radio, infrared, and optical polarimetry -- to analyze the data from X-ray polarimeters such as scattering polarimeters and photoelectric effect polarimeters, which measure the linear polarization of the detected X-rays.
Based on the azimuthal scattering angle (in the case of a scattering polarimeter) or the azimuthal component of the angle of the electron ejection (in the case of a photoelectric effect polarimeter), the Stokes parameters can be calculated for each event recorded in the detector. 
Owing to the additive nature of Stokes parameters, the analysis reduces to adding the Stokes parameters of the individual events and subtracting the Stokes parameters characterizing the background (if present). 
The main strength of this kind of analysis is that the errors on the Stokes parameters can be computed easily and are well behaved -- in stark contrast of the errors on the polarization fraction and polarization direction.
We demonstrate the power of the Stokes analysis by deriving several useful formulae, e.\,g.\ the expected error on the polarization fraction and polarization direction for a detection of $N_\mathrm{S}$ signal and $N_\mathrm{BG}$ background events, the optimal observation times of the signal and background regions in the presence of non-negligible background contamination of the signal, and the minimum detectable polarization (MDP) that can be achieved when following this prescription.
\end{abstract}


\end{frontmatter}


\section{Introduction}\label{sec:introduction}
The measurement of the linear polarization of the X-rays from cosmic sources holds the promise to provide geometrical information about the innermost regions of the most extreme objects in the universe, black holes and neutron stars~\cite{krawczynski_h_2011b,lei_f_1997,weisskopf_mc_2006}.
These systems emit copious amounts of X-rays but are too small to be imaged with current technology.
Despite the scientific potential of X-ray polarimetry, only one dedicated satellite-borne X-ray polarimeter has been flown so far. 
The Bragg polarimeter on board the OSO-8 satellite launched in 1978 measured a polarization fraction of the Crab Nebula of about~$20\%$ at energies of~$2.6\un{keV}$ and~$5.2\un{keV}$~\cite{weisskopf_m_c_1978}.
Since then, three more X-ray polarization measurements have been published: In 2008, the instruments SPI and IBIS on board the INTEGRAL satellite reported polarization fractions of the Crab Nebula of~$46\pm10\%$~\cite{dean_aj_2008} and~${>}72\%$~\cite{forot_m_2008}, respectively, with the polarization direction aligned with the X-ray jet.
For the stellar mass black hole Cygnus X-1 in an X-ray binary, a polarization fraction of~$40\pm10\%$ in the $230$ to $400\un{keV}$ range and ${>}75\%$~\cite{jourdain_e_2012} and~$67\pm30\%$~\cite{laurent_p_2011} at higher energies have been reported.
For a number of Gamma-Ray Bursts, tentative evidence for polarized emission has been published~\cite{coburn_w_2003,kalemci_e_2007,yonetoku_d_2011}, but the measurements are plagued by large statistical and systematic uncertainties.

More recently, various wider-bandpass polarimeters have been developed, including photoelectric effect polarimeters (e.g.\ the polarimeters of the proposed GEMS~\cite{hill_je_2012} and XIPE~\cite{soffitta_p_2013} missions) and scattering polarimeters (e.g.\ the polarimeters of the X-Calibur~\cite{guo_q_2013} and PoGOLite~\cite{pearce_m_2012} missions).
Photoelectric effect polarimeters track the direction of photoelectrons which are preferentially emitted parallel to the electric field of the incoming photons (e.\,g.\ Ref.~\cite{heitler_w_1936}).
Scattering polarimeters measure the direction into which the photons scatter and make use of the fact that photons scatter preferentially perpendicular to the electric field direction of the X-ray beam (e.\,g.\ Ref.~\cite{evans_r_1955}).
Unlike radio or optical telescopes, which measure the intensity of the radiation from the source, most X-ray telescopes detect individual photons. 
The linear polarization of the X-rays leads to a sinusoidal modulation of the azimuth distribution of events with a $180\degree$ period and a phase depending on the polarization direction.
The relative amplitude of the modulation corresponds to the polarization fraction.
The standard method for determining the linear polarization fraction and angle of an X-ray beam is to fit a sine function to the observed azimuth distribution.

In 1852, George Gabriel Stokes introduced a set of four parameters which are sufficient to completely describe the polarization properties of a quasi-monochromatic beam with arbitrary linear and circular polarization properties \cite{stokes_gg_1852}. 
These four \emph{Stokes Parameters} are linear, i.\,e.\ the intensity and polarization of a superposition of light beams from two different sources is described by the sum of their Stokes Parameters. 
Radio antennas and optical telescopes equipped with polarization filters -- being sensitive to certain polarization directions -- can basically measure Stokes Parameters directly (e.\,g.\ Ref.~\cite{hamaker_jp_1996}).
Owing to their additive properties, Stokes parameters are also used in theoretical calculations, e.\,g.\ in radiative transfer calculations \cite{chandrasekhar_s_1960} and in quantum mechanical calculations involving polarized photons~\cite{mcmaster_wh_1954,mcmaster_wh_1961}.

In this paper, we discuss the use of Stokes parameters in the analysis of the data from X-ray polarimeters.
We define the Stokes parameters for an idealized polarimeter with uniform acceptance and discuss their statistical properties in Section~\ref{sec:stokes}.
In Section \ref{sec:polarization} we describe the implications for the analysis of X-ray polarimetry data.
We give a detailed discussion of how to calculate errors on the polarization fraction and polarization direction in Section \ref{sec:bayes}.
In Section~\ref{sec:backgrounds}, we use the results from the previous sections to optimize the observation strategy in the presence of non-negligible backgrounds.
Finally, in Section~\ref{sec:summary} we summarize our findings. 
In the~\ref{sec:appendixA}, we give the equations in modified form for the case of a polarimeter with non-uniform detector acceptance.

\section{The Stokes parameters and their statistical properties}\label{sec:stokes}
\begin{figure}
  \centering
  \includegraphics[width=\figurewidth]{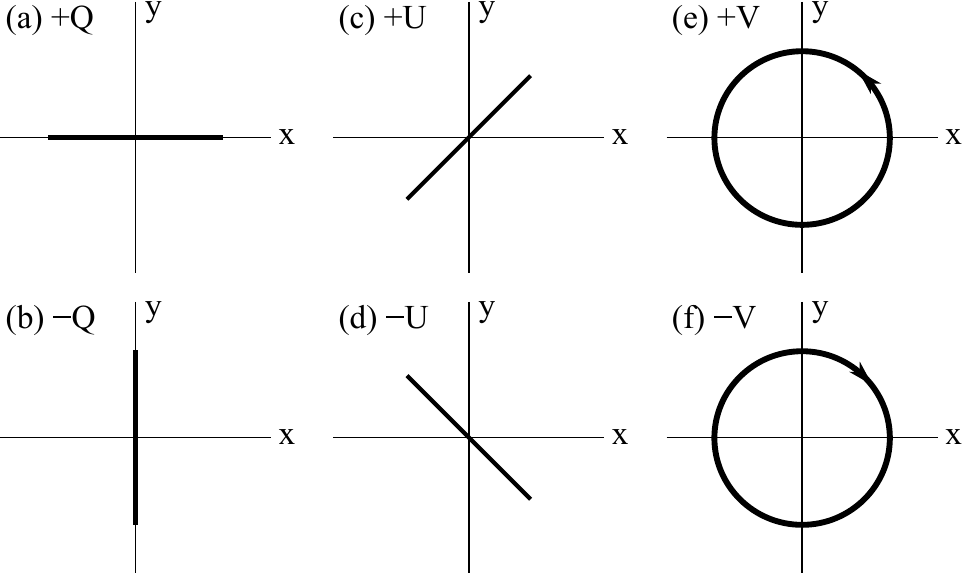}
  \caption{Polarization for different values of the Stokes parameters. (a)~$Q>0, U=0, V=0$; (b)~$Q<0, U=0, V=0$; (c)~$U>0, Q=0, V=0$; (d)~$U<0, Q=0, V=0$; (e)~$V>0, Q=0, U=0$; (f)~$V<0, Q=0, U=0$.}
  \label{fig:stokes-parameters}
\end{figure}

For a classical electromagnetic "quasi-monochromatic wave" (a wave which is 100\% polarized over short time intervals comparable to the period of the wave, but whose polarization properties change on longer time scales) the Stokes parameters can be defined by time averages (denoted by "$\langle\,\,\rangle$") of the electric field strength along two orthogonal directions (see e.\,g. Refs.~\cite{chandrasekhar_s_1960,rybicki_lightman_1986}).
Assuming a wave propagating along the $z$-axis towards larger $z$, the defining equations read:
\begin{subequations}\begin{align}
  \label{eq:stokes-I}S_0 &= I = \langle E_x^2 + E_y^2 \rangle,\\
  \label{eq:stokes-Q}S_1 &= Q = \langle E_x^2 - E_y^2 \rangle,\\
  \label{eq:stokes-U}S_2 &= U = \langle 2E_xE_y\cos\delta \rangle,\\
  \label{eq:stokes-V}S_3 &= V = \langle 2E_xE_y\sin\delta \rangle 
\end{align}\end{subequations}
Here, $E_x$ ($E_y$) is proportional to the instantaneous electric field along the x-axis (y-axis), 
$\delta$ is the lag of $E_y$ behind $E_x$. 
All four Stokes parameters have units of intensity (or flux). 
The parameter $I$ is the intensity (or flux) of the wave, $Q$ and $U$ depend on the linear polarization properties, and $V$ on the circular polarization properties.
The parameter $Q$ equals $I$ (-$I$) for a 100\% linearly polarized wave with an $\mathbf{E}$-field vector along the $x$-axis ($y$-axis); $U$ equals $I$ (-$I$) for a 100\% linearly polarized wave with an $\mathbf{E}$-field vector along the diagonal between the $x$-axis and the $y$-axis (the negative $x$-axis and the $y$-axis); $V$ equals $I$ (-$I$) for 100\% circularly right handed (left handed) polarized light.
Figure~\ref{fig:stokes-parameters} illustrates this.
The appropriate reference coordinate system for~$Q$ and~$U$ has been defined by the IAU~\cite{iau_vol15b}: $+Q$ corresponds to a linear polarization in North/South direction, $-Q$ to a polarization in East/West direction, and $+U$ corresponds to a polarization along the North East/South West diagonal.

For a monochromatic (100\% polarized wave) the identity 
\begin{equation}
  I\,=\,\sqrt{Q^2+U^2+V^2}
\end{equation}
holds. 
For a linearly polarized wave ($V=0$), the equation simplifies to $I\,=\,\sqrt{Q^2+U^2}$, and $Q$ and $U$ are given by the polarization direction $\psi$ (the angle between the $x$-axis and the electric field direction) by:
\begin{subequations}\begin{align}
  Q & = \cos{2\psi},      \\
  U & = \sin{2\psi},
\end{align}\end{subequations}
which implies:
\begin{equation}
  \tan{2 \psi} = \frac{U}{Q}.
\end{equation}

It can be shown that the Stokes parameters of an ensemble of quasi-monochromatic waves are additive: 
the Stokes parameters of the superposition of the waves equal the sum of the Stokes parameters of the individual waves. 
Such an ensemble of waves can be described as a superposition of unpolarized and polarized waves. 
It can be shown that the polarization fraction (the intensity or flux of the polarized waves) is then given by:
\begin{equation}
  p = \frac{\sqrt{Q^2+U^2+V^2}}{I}.
\end{equation}
In the case of linearly polarized waves ($V=0$), the linear polarization fraction is given by  
\begin{equation}\label{eq:linear_polarization_fraction}
  p_\mathrm{l} = \frac{\sqrt{Q^2+U^2}}{I}
\end{equation}
and the polarization direction (the direction of the electric field vector of the linearly polarized waves) can be inferred from the equation:
\begin{equation}
  \tan{2\psi} = \frac{U}{Q}.
\end{equation}

We now discuss how to use the Stokes parameters for the analysis of the data from an X-ray polarimeter, i.e. a scattering polarimeter or a photoelectric effect polarimeter. 
For both types of polarimeters, a data set consists of a list of $K$ angles $\left\{\varphi_k\right\}$ with $k=1 \ldots K$, which are related to the most likely azimuthal angle $\psi_k$ of the electric field vector.
In case of photoelectric effect polarimeters, $\psi_k = \varphi_k$, whereas in case of scattering polarimeters $\psi_k = \varphi_k - 90\degree$.
These angles exhibit a sinusoidal modulation with period $180\degree$~\cite{lei_f_1997,weisskopf_mc_2006,krawczynski_h_2011b}:
\begin{equation}\label{eq:azimuth_distribution}
  f(\psi) = \frac{1}{2\pi}\bigl(1 + p_0 \, \mu \cos(2 (\psi - \psi_0))\bigr),
\end{equation}
with $p_0$ being the true polarization fraction, $\psi_0$ giving the expected direction where the $\psi$-distribution peaks, and $\mu$ being the modulation factor. 
The modulation factor depends on the physics of the interaction and on the properties of the polarimeter and is defined as the amplitude of the azimuthal modulation measured for a $100\%$ polarized beam (i.\,e.\ for $p_0=1$), thus $0 \leq \mu \leq 1$.

For each event, we define a set of Stokes Parameters:
\begin{subequations}\label{eqs:single-stokes}\begin{align}
  \label{eq:single-stokes-i}i_k &= 1,           \\
  \label{eq:single-stokes-q}q_k &= \cos2\psi_k, \\
  \label{eq:single-stokes-u}u_k &= \sin2\psi_k,
\end{align}\end{subequations}
where we omitted an expression for $v_k$ as the polarimeter does not constrain the circular polarization. 
The Stokes parameters of the entire data set are then given by:
\begin{subequations}\label{eqs:total-stokes}\begin{align}
  \label{eq:total-stokes-I}I &= \sum_{k=1}^N i_k = N, \\
  \label{eq:total-stokes-Q}Q &= \sum_{k=1}^N q_k, \\
  \label{eq:total-stokes-U}U &= \sum_{k=1}^N u_k.
\end{align}\end{subequations}
It is convenient to introduce normalized Stokes parameters:
\begin{subequations}\label{eqs:total-stokes2}\begin{align}
  \label{eq:total-stokes-Q2}\QQ &= Q/I, \\
  \label{eq:total-stokes-U2}\UU &= U/I.
\end{align}\end{subequations}

For the sinusoidally modulated $\psi$-distribution of Equation (\ref{eq:azimuth_distribution}), we can calculate the expected $\QQ$ and $\UU$ values:
\begin{subequations}\label{eqs:qu_expectation}\begin{align}
  \label{eq:q_expectation}\langle \QQ \rangle &= \int\limits_0^{2\pi}\cos(2\psi)f(\psi) \, \dd\psi \,= \,\frac{1}{2} p_0 \, \mu \cos(2 \psi_0), \\
  \label{eq:u_expectation}\langle \UU \rangle &= \int\limits_0^{2\pi}\sin(2\psi)f(\psi) \, \dd\psi\,=\,\frac{1}{2} p_0 \, \mu \sin(2 \psi_0).
\end{align}\end{subequations}
We thus infer:
\begin{equation}\label{eq:q2+u2}
  \langle \QQ \rangle^2 + \langle \UU \rangle^2 = \frac{p_0^{\,2}  \mu^2}{4},
\end{equation}
and
\begin{equation}\label{eq:tan2chi}
  \frac{\langle\UU\rangle}{\langle\QQ\rangle} = \tan 2\psi_0.
\end{equation}
In analogy to Equations~(\ref{eqs:qu_expectation}a) and~(b) we can calculate the expected variance of the $q$-values.
After some algebra, we obtain:
\begin{equation}\begin{split}
  \bigl\langle(q-\langle q\rangle)^2\bigr\rangle &= \int\limits_0^{2\pi}\left(\cos(2\psi) - \frac{p_0\,\mu}{2} \cos(2 \psi_0)\right)^2 f(\psi) \, \dd\psi \\
  &= \frac{1}{2} -\frac{\mu^2 p_0^2}{4}\cos^2(2 \psi_0).
\end{split}\end{equation}
In the same way, we get:
\begin{equation}
  \bigl\langle (u-\langle u\rangle)^2 \bigr\rangle = \frac{1}{2} -\frac{\mu^2 p_0^2}{4}\sin^2(2 \psi_0).
\end{equation}
Thus, under the null hypothesis of no polarization ($p_0=0$), the variance of $q$ and $u$ are maximal and both equal~$1/2$.
We will use this result below to estimate the statistical significance of the detection of a polarized signal.
For a 100\% polarized signal and a hypothetical polarimeter with $\mu=1$, the variance of $q$ or $u$ can
be half of the maximum value. 
The root mean squared (RMS) daviations of $\QQ$ and $\UU$ from their average values are thus:
\begin{subequations}\label{eqs:rms}\begin{align}
  \label{eq:rms_q}\RMS(\QQ) &= \sqrt{\frac{1}{N} \left(\frac{1}{2} -\frac{\mu^2 p_0^2}{4}\cos^2(2 \psi_0)\right)}, \\
  \label{eq:rms_u}\RMS(\UU) &= \sqrt{\frac{1}{N} \left(\frac{1}{2} -\frac{\mu^2 p_0^2}{4}\sin^2(2 \psi_0)\right)}.
\end{align}\end{subequations}

\begin{figure}
  \centering
  \includegraphics[width=\figurewidth]{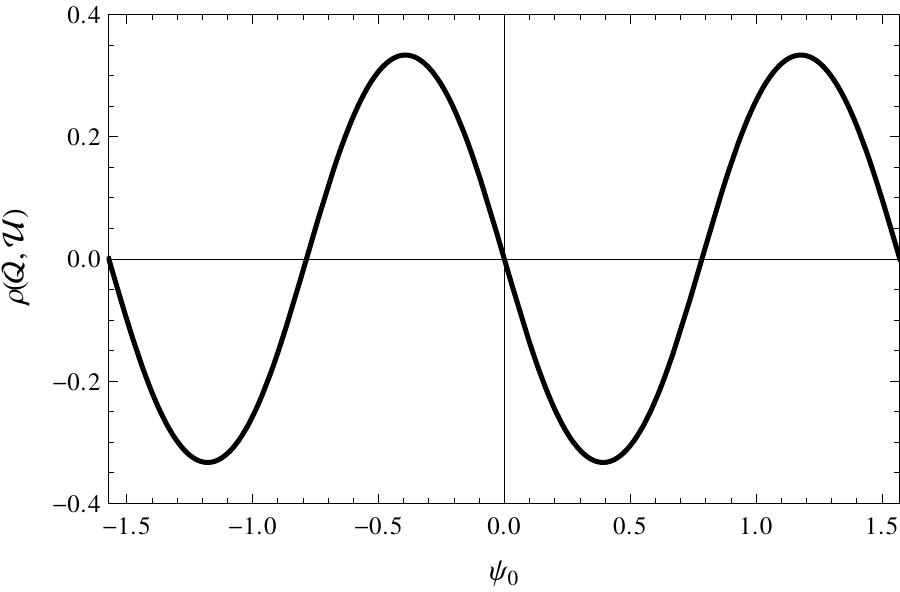}
  \caption{Correlation of the Stokes Parameters $\QQ$ and $\UU$ as a function of true polarization angle $\psi_0$, assuming a polarization fraction and modulation factor of $p_0=\mu=1$.}
  \label{fig:QU_correlation_chi}
\end{figure}

Along the same lines, we obtain the covariance of $\QQ$ and $\UU$:
\begin{equation}\label{eq:covariance}
  \cov(\QQ, \UU) = -\frac{1}{N} \frac{p_0^2\mu^2}{8}\sin(4\psi_0),
\end{equation}
and the linear correlation coefficient:
\begin{equation}\label{eq:correlation}
  \rho(\QQ, \UU) = -\frac{p_0^2\mu^2 \sin(4\psi_0)}{\sqrt{16 - 8p_0^2\mu^2 + p_0^4\mu^4\sin^2(4\psi_0)}}.
\end{equation}
Note that the correlation vanishes for the case of zero polarization ($p_0=0$).  
The correlation depends on the true angle owing to the coordinate frame dependence of $\QQ$ and $\UU$ (Fig.~\ref{fig:QU_correlation_chi}). 
Even for non-zero polarization, the correlation vanishes whenever $\psi_0$ is a multiple of $45\degree$, and thus when $\langle\QQ\rangle = 0$ or $\langle\UU\rangle = 0$. 
The maximum value for $\psi_0 = -22.5\degree$ is 
\begin{equation}
  \rho_\text{max} = \frac{1}{3}p_0^2\mu^2.
\end{equation}

\section{Reconstruction of the Polarization Fraction and Polarization Direction}\label{sec:polarization}
Given the results from a measurement $\left\{\psi_k\right\}$ with $k=1 \ldots K$, the Stokes parameters $\QQ$ and $\UU$ can be calculated according to Equations~(\ref{eqs:total-stokes2}a) and~(b).
From the expectation values in Equations~\eqref{eq:q2+u2} and~\eqref{eq:tan2chi} it follows that we can calculate the reconstructed polarization fraction $p_r$ and position angle $\psi_r$ from the measurement of $\QQ$ and $\UU$:
\begin{equation}\label{eq:measured-p}
  p_r = \frac{2}{\mu} \sqrt{\QQ^2 +\UU^2},
\end{equation}
and
\begin{equation}\label{eq:measured-chi}
  \tan2\psi_r = \frac\UU\QQ \quad\text{or}\quad \psi_r = \frac{1}{2}\arctan\frac\UU\QQ.
\end{equation}
These values are the best estimate of the true values $p_0$ and $\psi_0$ given the measurement of $\QQ$ and $\UU$.
The factor $2/\mu$ in Eq.~\eqref{eq:measured-p} results from the fact that the observed Stokes parameters are influenced by the modulation factor of the instrument, and that they are derived from the scattering angle distribution of photons or the emission angle distribution of photo electrons.
These angles follow a sinusoidal distribution with a~$180\degree$ period and minima/maxima corresponding to the polarization angle of the incident photon, which reduces the Stokes parameters derived from them by a factor~$1/2$ compared to the true Stokes parameters of the incident photons.
Furthermore, the observed Stokes parameters are reduced by a factor~$\mu$ due to the modulation factor of the instrument.
This can be seen in equations~\eqref{eq:q_expectation} and~(b).

How significant is the detection of a non-zero polarization? 
As noted above, $\QQ$ and $\UU$ are uncorrelated for the null hypothesis of zero polarization. 
According to the central limit theorem, $\QQ$ and $\UU$ are thus normally distributed for $N \gg 1$ with a mean of $0$ and Gaussian widths of (cf.\ Equations~(\ref{eqs:rms}a) and~(b)):
\begin{equation}
  \sigma_\QQ = \sigma_\UU\,=\,\frac{1}{\sqrt{2N}}.
\end{equation}
Under the null hypothesis, the probability density function for $\QQ$ and $\UU$ is thus given by
\begin{equation}
  f(\QQ, \UU) \, \dd\QQ \, \dd\UU = \frac{N}{\pi}e^{-N(\QQ^2+\UU^2)} \, \dd\QQ \, \dd\UU.
\end{equation}
Thus, if the detection of $N$ events gives the polarization fraction $p_r$ from Equation (\ref{eq:measured-p}), the probability $P_c$ to find a larger apparent polarization by pure chance is given by integrating $f(\QQ,\UU)$ over all $\QQ^2+\UU^2$-values larger than the observed ones.
We get the result:
\begin{equation}\label{eq:p_prob}
  P_c = \exp\left(-\frac{N}{4}\mu^2p^2_r\right).
\end{equation}
The formalism allows us to calculate the minimum detectable polarization (MDP) on 99\% confidence level
as function of $N$ and $\mu$ by solving Eq.~\eqref{eq:p_prob} for $p_r$, assuming $P_c = 1\%$:
\begin{equation}\label{eq:mdp}
  \text{MDP} \approx \frac{4.29}{\mu\,\sqrt{N}},
\end{equation}
which is exactly what previous authors have found (e.\,g. \cite{weisskopf_mc_2006,krawczynski_h_2011a}).
It is noteworthy that $\text{MDP} \propto \mu^{-1}$, which emphasizes the importance of optimizing the modulation factor when designing a polarimeter.
  
As discussed above the observed Stokes parameters~$\QQ$ and~$\UU$ are influenced by the modulation factor of the instrument, and they are derived from scattering or photo-electron emission angles, which themselves follow a sinusoidal distribution related to the polarization fraction and position angle.
These two effects reduce the observed Stokes parameters by a factor $\mu/2$ with respect to the true Stokes parameters of the incident beam.
Thus, an observer should quote the following reconstructed Stokes parameters, in order to make results from different experiments and theoretical calculations comparable:
\begin{subequations}\label{eqs:measured-stokes}\begin{align}
  \label{eq:measured-q}\QQ_r &= \frac{2}{\mu}\QQ, \\
  \label{eq:measured-u}\UU_r &= \frac{2}{\mu}\UU.
\end{align}\end{subequations}
Then, $p_r = \sqrt{\QQ_r^2 + \UU_r^2}$ holds as in the definition of the linear polarization fraction in Eq.~\eqref{eq:linear_polarization_fraction}.
The standard devitions of the parameters are:
\begin{subequations}\label{eqs:measured-sigma}\begin{align}
  \label{eq:measured-sigma_q}\begin{split}
    \sigma(\QQ_r) &= \frac{2}{\mu}\sqrt{\frac{1}{N-1}\left(\frac{1}{2} -\frac{\mu^2 p_r^2}{4}\cos^2(2 \psi_r)\right)} \\
      &= \sqrt{\frac{1}{N-1}\left(\frac{2}{\mu^2} - \QQ_r^2\right)},
  \end{split} \\
  \label{eq:measured-sigma_u}\begin{split}
    \sigma(\UU_r) &= \frac{2}{\mu}\sqrt{\frac{1}{N-1}\left(\frac{1}{2} -\frac{\mu^2 p_r^2}{4}\sin^2(2 \psi_r)\right)} \\
      &= \sqrt{\frac{1}{N-1}\left(\frac{2}{\mu^2} - \UU_r^2\right)},
  \end{split}
\end{align}\end{subequations}
and their covariance is given by:
\begin{equation} \label{eq:measured-cov_qu}
  \cov(\QQ_r, \UU_r) = -\frac{p_r^2}{2(N-1)} \sin(4\psi_r) = -\frac{\QQ_r \UU_r}{N-1}.
\end{equation}

Whereas the post-measurement probability distributions of $\QQ_r$ and $\UU_r$ are Gaussian distributions 
with well defined mean values and covariances, the post-measurement probability distributions of
polarization degree and polarization direction are not such simple distributions. 
We discuss those in the next section.

\section{Uncertainties on the Polarization Fraction and Polarization Direction}\label{sec:bayes}
A full analysis of the uncertainty intervals on the polarization fraction and polarization direction uses Bayes' theorem to infer the post-measurement probability distributions from the pre-measurement distributions~\cite{quinn_jl_2012}. 
Usually, one assumes that $Q$ and $U$ are normally distributed and uncorrelated~\citep{vinokur_m_1965,simmons_stewart_1985,vaillancourt_je_2006, weisskopf_mc_2006,krawczynski_h_2011a}:
\begin{multline}
  P(Q,U)=\\
    \frac{1}{2\pi\sigma^2}\exp\left[-\frac{(Q-\langle Q\rangle)^{2}+(U-\langle U\rangle)^{2}}{2\sigma^2}\right],
\end{multline}
or 
\begin{multline}
  P(p,\psi|p_0,\psi_0)=\\
    \frac{p}{2\pi\sigma^2}\exp\left[-\frac{p_0^2  + p^2 - 2pp_0\cos\bigl(2(\psi-\psi_0)\bigr)}{2\sigma^2}\right].
\end{multline}
The standard method to reconstruct polarization fraction and angle from X-ray polarimetry data is to fit the binned azimuthal distribution with a sine function.
Assuming Poissonian statistics and a total of $N$ events, one arrives at~\cite{weisskopf_mc_2006,krawczynski_h_2011a}:
\begin{multline}\label{eq:weisskopf_prob}
  P(p,\psi|p_0,\psi_0) = \frac{p \, \mu^2 N}{4\pi} \times \\
    \exp\left[-\frac{\mu^2 N}{4} \left(p^2 + p_0^2 - 2pp_0\cos\bigl(2(\psi-\psi_0)\bigr)\right)\right].
\end{multline}
Note that while this is the most commonly used method, binning in azimuth is not strictly necessary if an unbinned likelihood fit is performed instead.

We can derive an improved equation based on the results from Section \ref{sec:stokes}.
Assuming a bivariate normal distribution for $\QQ$ and $\UU$,
\begin{multline}
  P(\QQ, \UU) = \frac{1}{2\pi\sigma(\QQ)\sigma(\UU)\sqrt{1-\rho^2}} \times \\
    \exp\left[-\frac{1}{2(1-\rho^2(\QQ,\UU))}\left(\frac{(\QQ-\langle \QQ\rangle)^2}{\sigma^2(\QQ)} + \frac{(\UU-\langle \UU\rangle)^2}{\sigma^2(\UU)} \right.\right.\\
       \left.\left.-\frac{2\,\rho(\QQ,\UU)\,(\QQ - \langle \QQ\rangle)(\UU - \langle \UU\rangle)}{\sigma(\QQ)\sigma(\UU)}\right)\right],
\end{multline}
one finds for the polarization fraction and direction:
\begin{multline}\label{eq:pchi_prob}
  P(p,\psi|p_0,\psi_0) = \frac{\sqrt{N}\,p\,\mu^2}{2\pi\sigma} \times \\
    \exp\biggl[-\frac{\mu^2}{4\sigma^2} \biggl\{p_0^2+p^2 - 2pp_0\cos\bigl(2(\psi-\psi_0)\bigr) \\
    - \frac{p^2p_0^2\mu^2}{2}\sin^2\bigl(2(\psi-\psi_0)\bigr)\biggr\}\biggr]
\end{multline}
with
\begin{equation}\label{eq:pchi_sigma}
  \sigma = \sqrt{\frac{1}{N}\left(1-\frac{p_0^2\,\mu^2}{2}\right)}.
\end{equation}

This is an approximation since (a)~$\QQ^2 + \UU^2 \leq 1$ and, therefore, $\QQ$ and $\UU$ cannot be normally distributed for large values of $\QQ$ and $\UU$, and (b)~the correlation between $\QQ$ and $\UU$ is not exactly linear but follows a circle with radius~$p_0/2$. 
The former is not an issue because $\langle \QQ \rangle^2 + \langle \UU \rangle^2 \leq 1/4$.
With concern to the latter, $\sigma(\QQ),\sigma(\UU) \propto \sqrt{1/N} \ll 1$ for large $N$, so that the linear approximation of the correlation is good.

The most important difference between Eq.~\eqref{eq:pchi_prob} and Eq.~\eqref{eq:weisskopf_prob} is the third term in the exponent.
It accounts for the $p_0$-dependence of~$\QQ$ and~$\UU$, as well as the covariance of~$\QQ$ and~$\UU$.
Furthermore, $\sigma$ defined in Eq.~\eqref{eq:pchi_sigma} contains a second order correction in~$p_0$ accounting for the reduced variance at large~$p_0$.

At this point it is instructive to calculate a simple estimate of the uncertainty of the measurement of~$p_r$ and~$\psi_r$.
Using Equations~\eqref{eq:measured-p}--\eqref{eq:measured-chi} and~(\ref{eqs:measured-sigma}a,b)--\eqref{eq:measured-cov_qu} and standard error propagation one finds:
\begin{align}
  \label{eq:sigma-p_r}\sigma(p_r) &\approx \sqrt{\frac{2 - p_r^2\mu^2}{(N-1)\mu^2}}, \\
\intertext{and}
  \label{eq:sigma-psi_r}\sigma(\psi_r) &\approx \frac{1}{p_r\mu\sqrt{2(N-1)}}.
\end{align}
First, one should note that both $\sigma(p_r)$ and $\sigma(\psi_r) \propto \mu^{-1}$, again emphasizing the importance of a large modulation factor.
Additionally, as expected, the uncertainties are proportional to $(N-1)^{-1/2}$.
Furthermore, both uncertainties are smaller for highly polarized sources, but in general the effect of a large~$p_r$ is greater in~$\sigma(\psi_r)$.

By marginalizing the probability distribution in Equation~\eqref{eq:weisskopf_prob} over~$\psi$ and $p$ respectively, Weisskopf et al.\ (2006) find Gaussian approximations with~\cite{weisskopf_mc_2006}\footnote{Weisskopf et al.\ did not consider the modulation factor in their calculations. Here, the appropriate factors have been added for consistency. Furthermore, their expression for the polarization direction uncertainty differs by a factor $2$ because they use $\phi=2\psi$.}
\begin{align*}
  \sigma(p_r) &= \sqrt{2/(N\mu^2)}, \\
  \sigma(\psi_r) &= \sigma(p_r)/(2p_r).
\end{align*}
Equation~\eqref{eq:sigma-p_r} has an additional $p_r$ dependent correction, which results from the corresponding terms in Equations \eqref{eq:measured-sigma_q} and~\eqref{eq:measured-sigma_u}.
However, this correction will typically be much less than~$10\%$.

Using Bayes' theorem and $P(p, \psi | p_0, \psi_0)$ from Equation~\eqref{eq:pchi_prob}, as well as a prior distribution $P_0(p_0, \psi_0)$ one can now (numerically) compute the posterior distribution
\begin{equation}
  P(p_0, \psi_0 | p, \psi) = \frac{P(p, \psi | p_0, \psi_0) \, P_0(p_0, \psi_0)}{\int_0^\pi \int_0^1 P(p, \psi | p'_0, \psi'_0) \, P_0(p'_0, \psi'_0) \, \dd p'_0 \dd\psi'_0}.
\end{equation}
Different choices for the prior $P_0$ are discussed in Reference~\cite{quinn_jl_2012}.
If there is no prior knowledge about the polarization of a source, a good choice is $P_0 \equiv 1/\pi$, i.\,e.\ uniform in the $p_0$-$\psi_0$ plane.
This prior is not uniform in the $\QQ$-$\UU$ plane:
\begin{equation}
  P_0(\QQ, \UU) = \left(\pi\mu\sqrt{\QQ^2 + \UU^2}\right)^{-1},
\end{equation}
making this prior a subjective prior in a certain sense, prefering values closer to 0.
Its advantage over the Jeffreys prior, which is uniform in the $\QQ$-$\UU$ plane and an objective prior, is that it allows to infer the probability that a measurement is compatible with a zero polarization, and thus construction of an upper limit on the polarization fraction.

For a measurement of $p$ and $\psi$, contours for the confidence level $c$ can then be constructed by (numerically) solving 
\begin{equation}
  \iint\limits_{A(a)} P(p_0, \psi_0 | p, \psi) \, \dd p_0 \dd\psi_0 = c,
\end{equation}
to find a value of $a$, where $A(a)$ is the set of points for which $P(p_0, \psi_0 | p, \psi) > a$, $A(a) = \{(p_0,\psi_0); P(p_0, \psi_0 | p, \psi) > a\}$.
The confidence contour of level $c$ is then given by all points for which $P(p_0, \psi_0 | p, \psi) = a$.

\section{Experimental Backgrounds}\label{sec:backgrounds}
Stokes parameters can be used to describe not only the signal events, but also the backgrounds.
In an analysis of polarization data, one then uses a forward modeling analysis in which the sum of the model Stokes parameters and those of the background are compared to the measured Stokes parameters.
This works even if the backgrounds are anisotropic and thus mimic a polarization signal.
The background Stokes parameters can be obtained during special off-source observations observing a dark patch of the sky close to the targeted X-ray source.

In this section we discuss the best possible split between on-source and off-source observations which minimizes the statistical error on the resulting Stokes parameters.
To do this, we subtract the off-source Stokes parameters from the on-source Stokes parameters.
Observing the source for a duration of $t_\mathrm{on}$ and the background for a duration of $t_\mathrm{off}$, one weights the background $\QQ$ and $\UU$ parameters with the factor $w_\mathrm{off} = -\alpha^{-1}$ with
\begin{equation}
  \alpha = \frac{t_\mathrm{off}}{t_\mathrm{on}} = \frac{f_\mathrm{off}}{1-f_\mathrm{off}},
\end{equation}
where $f_\mathrm{off}$ is the fraction of the total observation time spent off source.
The negative sign of $w_\mathrm{off}$ results in a subtraction of the off-source events from the on-source data.
Weights of the on-source events will be $w_\mathrm{on} = 1$.
Analyzing the data from a polarimeter with non-uniform detector acceptance may require to weight the individual events with a weighting factor $w_k$ to account for example for dead space between different pixels (see~\ref{sec:appendixA}). 
We give the results of the optimization of $t_\mathrm{on}$ and $t_\mathrm{off}$ here for this more general case.

When preparing an observation, one possible way to determine the best value of~$\alpha$ is to minimize~$\sigma \propto \sqrt{W_2}/I$ defined in Eq.~\eqref{eq:pchi_sigma-weighted} with $W_2 = \sum w^2$ and $I=\sum w$.
Let $T$ be the total observation time, $R_\mathrm{S}$ the expected signal rate, and $R_\mathrm{BG}$ the expected background rate, then
\begin{equation}\label{eq:I_background}
  I = R_\mathrm{S} \, T \, (1-f_\mathrm{off})
\end{equation}
and
\begin{equation}\label{eq:W2_background}
  W_2 = (R_\mathrm{S} + R_\mathrm{BG}) \, T \, (1-f_\mathrm{off}) + R_\mathrm{BG} \, T \, f_\mathrm{off} \, \left(\frac{1-f_\mathrm{off}}{f_\mathrm{off}}\right)^2,
\end{equation}
where the first term corresponds to the on-source contribution and the second term to the off-source contribution.
Taking the derivative of $\sqrt{W_2}/I$ one finds the positive root
\begin{equation}\label{eq:foff}
  f_\mathrm{off} = \frac{\sqrt{R_\mathrm{BG}(R_\mathrm{BG}+R_\mathrm{S})}-R_\mathrm{BG}}{R_\mathrm{S}} = \frac{\sqrt{\smash[b]{1+R_\mathrm{S/B}}}-1}{R_\mathrm{S/B}},
\end{equation}
or
\begin{equation}
  \alpha = \frac{R_\mathrm{BG}}{\sqrt{R_\mathrm{BG}(R_\mathrm{BG}+R_\mathrm{S})}} = \frac{1}{\sqrt{\smash[b]{1+R_\mathrm{S/B}}}},
\end{equation}
where $R_\mathrm{S/B}=R_\mathrm{S}/R_\mathrm{BG}$ is the signal-to-background ratio.

Using these results, the minimum detectable polarization from Eq.~\eqref{eq:mdp} becomes (cf.\ Equation~\eqref{eq:mdp-weighted}):
\begin{align}
  \notag\text{MDP} &\approx \frac{4.29}{\mu \, I} \sqrt{W_2} = \frac{4.29}{\mu \, R_\mathrm{S}} \sqrt{\frac{R_\mathrm{BG} + f_\mathrm{off}R_\mathrm{S}}{(1-f_\mathrm{off})f_\mathrm{off}T}} \\
\intertext{and with Eq.~\eqref{eq:foff}}
  \text{MDP} &= \frac{4.29 \, \sqrt{\smash[b]{R_\mathrm{BG} + R_\mathrm{S}}}}{\mu \, \sqrt{T} \left(R_\mathrm{BG} + R_\mathrm{S} - \sqrt{R_\mathrm{BG}(R_\mathrm{BG}+R_\mathrm{S})}\right)}.
\end{align}
Note that we minimized~$\sigma$ in order to find the value of~$f_\mathrm{off}$ used here.
This also optimizes the MDP.

Using Equations~\eqref{eq:I_background}--\eqref{eq:foff}, the approximate errors on $p_r$ and $\psi_r$ from Equations~\eqref{eq:sigma-p_r} and~\eqref{eq:sigma-psi_r} become:
\begin{align}
  \notag\sigma(p_r) &= \sqrt{\frac{W_2(2-p_r^2\mu^2)}{I^2\mu^2}} \\
    &= \left[\frac{\rho_\mathrm{BS}(2-p_r^2\mu^2)}{\bigl(\rho_\mathrm{BS}(2R_\mathrm{BG} + R_\mathrm{S}) - 2(R_\mathrm{BG}^2 + R_\mathrm{BG}R_\mathrm{S})\bigr) \, T \, \mu^2}\right]^{\frac{1}{2}} \\
  \sigma(\psi_r) &= \frac{\sqrt{R_\mathrm{BG} + R_\mathrm{S}/2 + \rho_\mathrm{BS}}}{p_r \, R_\mathrm{S} \, \mu\,\sqrt{T}}
\end{align}
with $\rho_\mathrm{BS} = \sqrt{R_\mathrm{BG}(R_\mathrm{BG}+R_\mathrm{S})}$.

\section{Summary}\label{sec:summary}
Stokes parameters are a set of four parameters that fully describe the polarization properties of an electromagnetic wave.
They are a tool commonly used to analyze radio, infrared, and optical polarimetry data, where they can be measured directly.
The main advantages of the use of Stokes parameters over polarization fraction and direction are that they are additive and that their distribution is very well described by a bivariate normal distribution.

Unlike radio, infrared, and optical telescopes, X-ray instruments do not measure intensities, but detect individual photons.
Scattering polarimeters measure the azimuthal scattering angle of incident photons, and photoelectric effect polarimeters measure the angle of the photoelectron emission.
Based on this angle the Stokes parameters~$q$ and~$u$, which describe linear polarization, can be assigned to each event.
Thanks to the additive nature of the Stokes parameters, the linear polarization fraction and direction can then be calculated from the sum of the Stokes parameters of the individual events.
If there is a non-negligible background, its Stokes parameters can simply be subtracted from the observation.
The introduction of event weights in case on-source and off-source observation times differ is straightforward.

Furthermore, the Stokes parameters method avoids the information loss associated with the azimuthal binning of events commonly used when fitting the azimuthal distribution.
An alternative method to the explicit use of Stokes parameters, which also avoids binning, is an unbinned likelihood fit of the azimuth distribution as described, for example, in Ref.~\cite{krawczynski_h_2011a}.

An observer should not only quote the measured polarization fraction and direction, but also the normalized Stokes parameters~$\QQ$ and~$\UU$, and their uncertainties.
When doing so, authors should adhere to the convention for Stokes parameters set forth by the IAU~\cite{iau_vol15b}.
This will simplify comparison with theoretical results, since these are often given in terms of~$\QQ$ and~$\UU$.

The results presented here can also be used to fit models of the X-ray intensity and polarization properties to experimental data.
In this case, a model depending on a set of fitting  parameters ${\cal P}$ is used to predict the $\QQ_i$ and $\UU_i$ ($i = 1 \ldots M$) parameters in $M$ energy or temporal bins.
The parameters $\QQ_i$ and $\UU_i$ include the background Stokes parameters.
Equations~(\ref{eqs:measured-sigma}a, b) can be used to calculate the expected errors and Equation~\eqref{eq:covariance} the expected covariance of $\QQ_i$ and $\UU_i$.
The model parameters can then be fitted by minimizing the $\chi^2$ value
\begin{equation}\begin{split}
  \chi^2 = \sum_{i=1}^M \biggl[&\frac{1}{\sigma^2(\QQ_i)\sigma^2(\UU_i) - \cov^2(\QQ_i, \UU_i)} \times \\
    &\Bigl(\sigma^2(\UU_i)(\QQ_i^\mathrm{m}-\QQ_i)^2 + \sigma^2(\QQ_i)(\UU_i^\mathrm{m}-\UU_i)^2 \\
    &-2\cov(\QQ_i,\UU_i)(\QQ_i^\mathrm{m}-\QQ_i)(\UU_i^\mathrm{m}-\UU_i)\Bigr) \biggr],
\end{split}\end{equation}
where $\QQ_i^\mathrm{m}$ and $\UU_i^\mathrm{m}$ are the measured values in the $i$-th energy bin.
The model that is used to predict the $\QQ_i$ and $\UU_i$ may need to incorporate the energy response of the polarimeter, in particular if the energy resolution or asymmetries in the response are not negligible compared to the size of the energy bins.

In case of energy-resolved polarimetry, unfolding can be used as an alternative to the forward folding described above.
It has the advantage that one obtains model-in\-de\-pen\-dent data points of physical quantities.
A detailed description of such a method can be found in Ref.~\cite{kislat_f_2014}.
Stokes parameters can be introduced as the output of the unfolding algorithm, instead of the scattering angle as described in the reference.
However, in that case a binning of the Stokes parameter distributions will occur.

Because the probability distributions of the Stokes parameters are well-behaved Gaussians, energy dependent polarization models can easily be fit to the data by forward folding them into binned distributions of $\QQ$, $\UU$, and energy.
Model parameters can then be determined through a simple chi-squared minimization.

In conclusion, thanks to their additivity and well-be\-haved probability distributions, Stokes parameters are a useful tool for the analysis of data from X-ray polarimeters.

\appendix
\section{Equations for Polarimeters with Non-Uniform Acceptance}\label{sec:appendixA}
In case of non-uniform acceptance, events need to be weighted in order to correct for this non-uniformity.
Note, however, that gaps in the azimuthal coverage cannot be recovered by weighting events.
Similarly, as described in Section~\ref{sec:backgrounds}, weights need to be applied when subtracting background events in case of differing on-source and off-source observation times.

Weights are introduced by modifying the definitions in Equations (\ref{eqs:single-stokes}a--c):
\begin{subequations}\label{eqs:single-stokes-weighted}\begin{align}
  \label{eq:single-stokes-i-weighted}i_k &\to w_k,    \\
  \label{eq:single-stokes-q-weighted}q_k &\to w_kq_k, \\
  \label{eq:single-stokes-u-weighted}u_k &\to w_ku_k.
\end{align}\end{subequations}
Equations~(\ref{eqs:total-stokes}a-c) then read
\begin{subequations}\label{eqs:total-stokes-weighted}\begin{align}
  \label{eq:total-stokes-I-weighted}I &= \sum_{k=1}^Nw_k, \\
  \label{eq:total-stokes-Q-weighted}Q &= \sum_{k=1}^Nw_kq_k, \\
  \label{eq:total-stokes-U-weighted}U &= \sum_{k=1}^Nw_ku_k.
\end{align}\end{subequations}

Introducing
\begin{equation}\label{eq:w2}
  W_2 = \sum_{k=1}^Nw_k^2,
\end{equation}
Equations~(\ref{eqs:rms}a) and (b) now read:
\begin{subequations}\label{eqs:rms-weighted}\begin{align}
  \label{eq:rms_q-weighted}RMS(\QQ) &= \frac{W_2}{I^2} \left(\frac{1}{2} -\frac{\mu^2 p_0^2}{4}\cos^2(2 \psi_0)\right), \\
  \label{eq:rms_u-weighted}RMS(\UU) &= \frac{W_2}{I^2} \left(\frac{1}{2} -\frac{\mu^2 p_0^2}{4}\sin^2(2 \psi_0)\right);
\end{align}\end{subequations}
and the covariance from Equation~\eqref{eq:covariance} is
\begin{equation}\label{eq:covariance-weighted}
  \cov(\QQ, \UU) = -\frac{W_2}{I^2} \frac{p_0^2\mu^2}{8}\sin(4\psi_0),
\end{equation}
whereas the factor $W_2/I^2$ cancels out in the correlation coefficient $\rho(\QQ, \UU)$ and Equation~\eqref{eq:correlation} remains unchanged.

Using these results, the probability distribution for the polarization fraction and direction from Equation~\eqref{eq:pchi_prob} becomes:
\begin{multline}
   P(p,\psi|p_0,\psi_0) = \frac{\sqrt{I^2/W_2}\,p\,\mu^2}{2\pi\sigma} \times \\
    \exp\biggl[-\frac{\mu^2}{4\sigma^2} \biggl\{p_0^2+p^2 - 2pp_0\cos\bigl(2(\psi-\psi_0)\bigr) \\
    - \frac{p^2p_0^2\mu^2}{2}\sin^2\bigl(2(\psi-\psi_0)\bigr)\biggr\}\biggr],
\end{multline}
with
\begin{equation}\label{eq:pchi_sigma-weighted}
  \sigma = \sqrt{\frac{W_2}{I^2}\left(1 - \frac{p_0^2\mu^2}{2}\right)}.
\end{equation}
The MDP from Equation~\eqref{eq:mdp} reads:
\begin{equation}\label{eq:mdp-weighted}
  \text{MDP} \approx \frac{4.29}{\mu \, I} \sqrt{W_2}.
\end{equation}

For large $N$ (which is true for any useful polarization measurement), the standard deviations on the measured values of $\QQ_r$ and $\UU_r$ are:
\begin{subequations}\label{eq:measured-sigma-weighted}\begin{align}
  \label{eq:measured-sigma_q-weighted}\begin{split}    
    \sigma(\QQ_r) &= \sqrt{\frac{W_2}{I^2}\left(\frac{1}{2} -\frac{\mu^2 p_r^2}{4}\cos^2(2 \psi_r)\right)} \\
      &= \sqrt{\frac{W_2}{I^2}\left(\frac{2}{\mu^2}-\QQ_r^2\right)},
  \end{split} \\
  \label{eq:measured-sigma_u-weighted}\begin{split}
    \sigma(\UU_r) &= \sqrt{\frac{W_2}{I^2}\left(\frac{1}{2} -\frac{\mu^2 p_r^2}{4}\sin^2(2 \psi_r)\right)} \\
      &= \sqrt{\frac{W_2}{I^2}\left(\frac{2}{\mu^2}-\UU_r^2\right)},
  \end{split}
\end{align}\end{subequations}
and the covariance is:
\begin{equation}
  \label{eq:measured-cov_qu-weighted}\cov(\QQ_r, \UU_r) = -\frac{W_2}{I^2} \frac{p_r^2\mu^2}{8}\sin(4\psi_r) = -\frac{W_2}{I^2} \QQ_r\UU_r.
\end{equation}

\section*{Acknowledgements}
The authors are grateful for NASA funding from grants NNX10AJ56G \& NNX12AD51G as well as discretionary funding from the McDonnell Center for the Space Sciences.

\bibliography{references}
\bibliographystyle{elsarticle-num}

\end{document}